\documentclass[aps,prl,twocolumn,preprintnumbers,superscriptaddress,10pt]{revtex4-1}

\pdfoutput=1

\usepackage[T1]{fontenc} 
\usepackage{graphicx}

\usepackage[latin1]{inputenc}

\usepackage{color}

\begin{document}

\preprint{ITP-UU-13/22}

\title{A fully dynamical simulation of central nuclear collisions}

\author{Wilke van der Schee}
\affiliation{Institute for Theoretical Physics and Institute for Subatomic Physics,
Utrecht University, Leuvenlaan 4, 3584 CE Utrecht, The Netherlands}
\author{Paul Romatschke}
\affiliation{Department of Physics, 390 UCB, University of Colorado, Boulder, CO 80309-0390, USA}
\author{Scott Pratt}
\affiliation{Department of Physics and Astronomy and National Superconducting Cyclotron Laboratory,
Michigan State University, East Lansing, Michigan 48824, USA}


\begin{abstract}
We present a fully dynamical simulation of central nuclear collisions around mid-rapidity at LHC energies. Unlike previous treatments, we simulate all phases of the collision, including the equilibration of the system. For the simulation, we use numerical relativity solutions to AdS/CFT for the pre-equilibrium stage, viscous hydrodynamics for the plasma equilibrium stage and kinetic theory for the low density hadronic stage. Our pre-equilibrium stage provides initial conditions for hydrodynamics, resulting in sizable radial flow.
The resulting light particle spectra reproduce the measurements from the ALICE experiment at all transverse momenta.
%
\end{abstract}

\maketitle

\noindent
{{\bf 1. Introduction.}}
High precision experimental data from nucleus-nucleus collisions at the Relativistic Heavy Ion Collider (RHIC) and the Large Hadron Collider (LHC) have  led to a refined understanding of Quantum Chromodynamics (QCD) at high temperatures. According to the current paradigm, the colliding matter thermalizes fast, after which it expands hydrodynamically and finally hadronizes into a gas of particles. This ``fluid-like'' behavior is understood to arise from strongly coupled dynamics, which makes the quark-gluon plasma both interesting and complicated. Moreover, recent experimental data seems to be suggesting that fluid-like behavior is also seen in proton-nucleus and deuteron-nucleus collisions (cf.~\cite{Abelev:2012cya}). Also, it has been found that a particular observable (``elliptic flow'') is essentially insensitive to the collision energy (cf.~\cite{Adamczyk:2012ku}). These experimental findings indicate that a more refined theoretical understanding of high energy nuclear collisions is needed.

Simulations involving thermalization, hydrodynamics and hadronization can be compared to experimental data in an effort to extract properties of high density QCD matter, such as the viscosity coefficient $\eta$. These simulations, however, involve several unknown parameters and functions, such as the initial conditions for hydrodynamics (starting time $\tau_{\rm hydro}$, energy density, velocity, shear tensor). Understanding these unknown parameters can therefore lead to more precise extraction of QCD properties, and consequently to a better understanding of strongly coupled theories like QCD.

In particular, very little is known about the far-from-equilibrium evolution towards hydrodynamics, such that until now reasonable initial conditions at  $\tau_{\rm hydro}$ had to be guessed. In this Letter we attempt to reduce this ambiguity by including the far-from-equilibrium stage obtained from colliding shock waves in AdS, which through the AdS/CFT correspondence are thought to be similar to heavy ion collisions. After the matter has equilibrated, we match the AdS/CFT results onto a standard viscous hydrodynamics code which, once the matter has cooled below the QCD phase transition temperature $T_c$, is itself matched onto a standard hadronic cascade code, thereby achieving a fully dynamical simulation of a boost-invariant heavy-ion collision.

\noindent
{{\bf 2. Methodology.}}
The main physical input for our simulation will be the energy density of a highly boosted  and Lorentz contracted nucleus, $T_{tt} = \delta(t+z) T_A(x,\,y)$, with the ``thickness function''
\begin{equation}
\label{Fermi}
T_A(x,\,y) =  \epsilon_0 \int_{-\infty}^\infty dz \left[1+e^{(\sqrt{x^2+y^2+z^2}-R)/a}\right]^{-1}\,,
\end{equation}
where  $R=6.62$ fm, $a=0.546$ fm for a $^{208}{\rm Pb}$ nucleus\cite{Alver:2008aq}. The normalization $\epsilon_0$ is a measure of the energy of the nucleus, 
and given the simplicity of our model, we will use this constant to match the experimentally observed number of particles (``multiplicity'', $dN/dY$).

The AdS/CFT correspondence gives the dynamics of the stress-energy tensor $T^{\mu\nu}$ of the boundary strongly coupled CFT using the gravitational field in the bulk of $AdS_5$. A relativistic nucleus may therefore be described using a gravitational shockwave in AdS, whereby the stress-energy tensor of a nucleus can be exactly matched \cite{deHaro:2000xn}. In this work we limit ourselves to pure gravity, which is the dominant force at high energies, but generalizations are straightforward. For a head-on (central) collision this shockwave collision has been written down and solved near the boundary of AdS in Ref.~\cite{Romatschke:2013re}, resulting in the stress-energy tensor at early times. In polar Milne coordinates $\tau,\xi,\rho,\theta$ with $t=\tau \cosh \xi$, $z=\tau \sinh \xi$, $\rho^2=x^2+y^2$, $\tan\theta=y/x$,  $T^{\mu\nu}$ can be decomposed into
\begin{equation}
\label{Tmunuhydro}
T^{\mu\nu}=(\epsilon+P(\epsilon))u^\mu u^\nu + P(\epsilon) g^{\mu\nu}+\pi^{\mu\nu}\,,
\end{equation}
with $\epsilon,P(\epsilon),u^\mu,\pi^{\mu\nu}$ the local energy density, equation of state (EoS), velocity and shear tensor. In Ref.~\cite{Romatschke:2013re}, it was found to leading order in $t$ that
\begin{eqnarray}
\label{earlytimeTmunu}
\epsilon= 2 T_A^2(\rho)\tau^2\,,\ u^{\rho}=-\frac{T_A^\prime(\rho)}{3 T_A(\rho)} \tau\,,\ \frac{P_L}{P_T}=-\frac{3}{2},\quad
\end{eqnarray}
where in the local rest frame $T^\mu_\nu={\rm diag}(-\epsilon,P_T,P_T,P_L)$. The velocity dependence and pressure anisotropy are consistent with previous results ~\cite{Vredevoogd:2008id,Casalderrey-Solana:2013aba,Grumiller:2008va,Taliotis:2010pi}. 
To leading order in $\tau$, the corresponding line-element $ds^2$ turns out to be $\xi$-independent (boost-invariant) and can be transformed into the form
\begin{eqnarray}
ds^2&=&-A d\tau^2+\Sigma^2 \left(e^{-B-C} d\xi^2+e^{B}d\rho^2+e^{C} d\theta^2\right)
\nonumber\\
&&+2 dr d\tau+2 F d\rho d\tau\,,
\end{eqnarray}
where all functions depend on $\tau$, $\rho$ and the fifth AdS spacetime dimension $r$ only. The space boundary is located at $r\rightarrow \infty$ where the induced metric becomes $g_{\mu\nu}={\rm diag}(g_{\tau\tau},g_{\rho\rho},g_{\theta\theta},g_{\xi\xi})={\rm diag}(-1,1,\rho^2,\tau^2)$, which has coordinate singularities at $\tau=0$ and $\rho=0$. The metric can be expanded near the boundary, e.g. $B(r,\tau,\rho)=B_0(r,\tau,\rho)+\sum_{i=0}^\infty b_i(\tau,\rho)r^{-i}$, with $B_0$  given by the vacuum value. The early time result (\ref{earlytimeTmunu}) fixes the first few near-boundary series coefficients, but does not fix the metric functions deep in the bulk, leading to an unstable time evolution. In order to have a stable time evolution, we introduce a  function with one bulk parameter $\sigma$ to extend the metric functions to arbitrary $r$, specifically choosing
\begin{equation}
\label{sigma}
B(r,\tau,\rho)\rightarrow B_0(r,\tau,\rho)+\sum_{i=0}^6  \frac{b_i(\tau,\rho)r^{-i}}{1+\sigma^7 r^{-7}}\,,
\end{equation}
and analogously for $C$. With the near-boundary coefficients fixed by Eq.~(\ref{earlytimeTmunu}) at a time $\tau_{\rm init}$ and choosing a value for $\sigma$, the future metric is completely determined and is obtained by numerically solving Einstein equations adopting 
a pseudo-spectral method based on \cite{Chesler:2010bi, vanderSchee:2012qj,Chesler:2013aa}\cite{footnote1} 

From the metric we can extract the full $T^{\mu\nu}$ and in particular observe the transition from early-time, far-from equilibrium dynamics to a fluid described by viscous hydrodynamics. At some value of proper time $\tau_{\rm hydro}$, we stop the evolution using Einstein equations and extract $\epsilon,u^\mu,\pi^{\mu\nu}$ from Eq.~(\ref{Tmunuhydro}). These functions provide the initial conditions for the well-tested relativistic viscous hydrodynamic code vh2 (version 1.0) \cite{Luzum:2008cw}, which uses an EoS inspired by lattice QCD 
and has, for simplicity, $\eta/s=\frac{1}{4\pi}$. Since this EoS differs from the conformal EoS of our AdS model there will be a discontinuity in the pressure. At high temperatures, however, QCD is approximately conformal and in our simulations the discontinuity at the center was never more than 15\%.

The hydrodynamic code simulates the evolution from $\tau=\tau_{\rm hydro}$ until the last fluid cell has cooled down below $T_{\rm sw}=0.17$ GeV. The hydrodynamic variables along the hypersurface defined by $T=T_{\rm sw}$ are stored and converted into particle spectra using the technique from Ref.~\cite{Pratt:2010jt}. The subsequent particle scattering is treated using a hadron cascade \cite{Novak:2013bqa} for resonances with masses up to $2.2$ GeV by simulating $500$ Monte-Carlo generated events. Once the particles have stopped interacting, and particles unstable under the strong force have decayed, light particle transverse momentum spectra are analyzed and can be compared to data.

From the hydrodynamic evolution onward our model uses techniques and parameters which are fairly standard. The initial conditions for hydrodynamics, however, are now determined using a far-from-equilibrium evolution. We modeled this phase as a strongly coupled CFT, described by gravity in AdS. This introduces new parameters and functions, namely the initialization time $\tau_{\rm init}$, the normalization $\epsilon_0$, the bulk function with parameter $\sigma$ and the AdS/hydro switching time $\tau_{\rm hydro}$. We will explore the effects of changing these parameters below.

\begin{figure}[t]
\includegraphics[width=\linewidth]{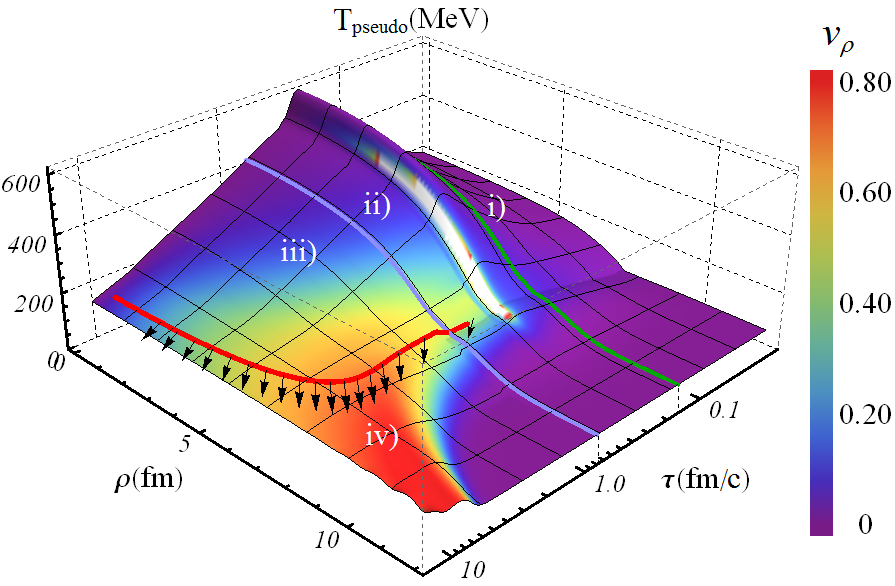}
\caption{Assuming (\ref{Tmunuhydro}) applied, a 
``pseudo'' temperature (defined by using Eq.~(\ref{Tmunuhydro}) with $\epsilon=\epsilon(T_{\rm pseudo})$) and radial velocity $v_\rho$ are extracted 
for a representative simulation.
The plot illustrates four physical tools used: i) early time expansion, ii) numerical AdS evolution, iii) viscous hydrodynamics until $T=0.17 \rm{GeV}$, iv) kinetic theory after 
conversion into particles (indicated by arrows). The (white) region close to $\tau\sim 0.2 $ fm/c, $\rho\sim 5$ fm indicates a far-from-equilibrium domain where a local rest frame cannot be found.}
\label{fig:Temp1}
\end{figure}

\begin{figure}[t]
\includegraphics[width=\linewidth]{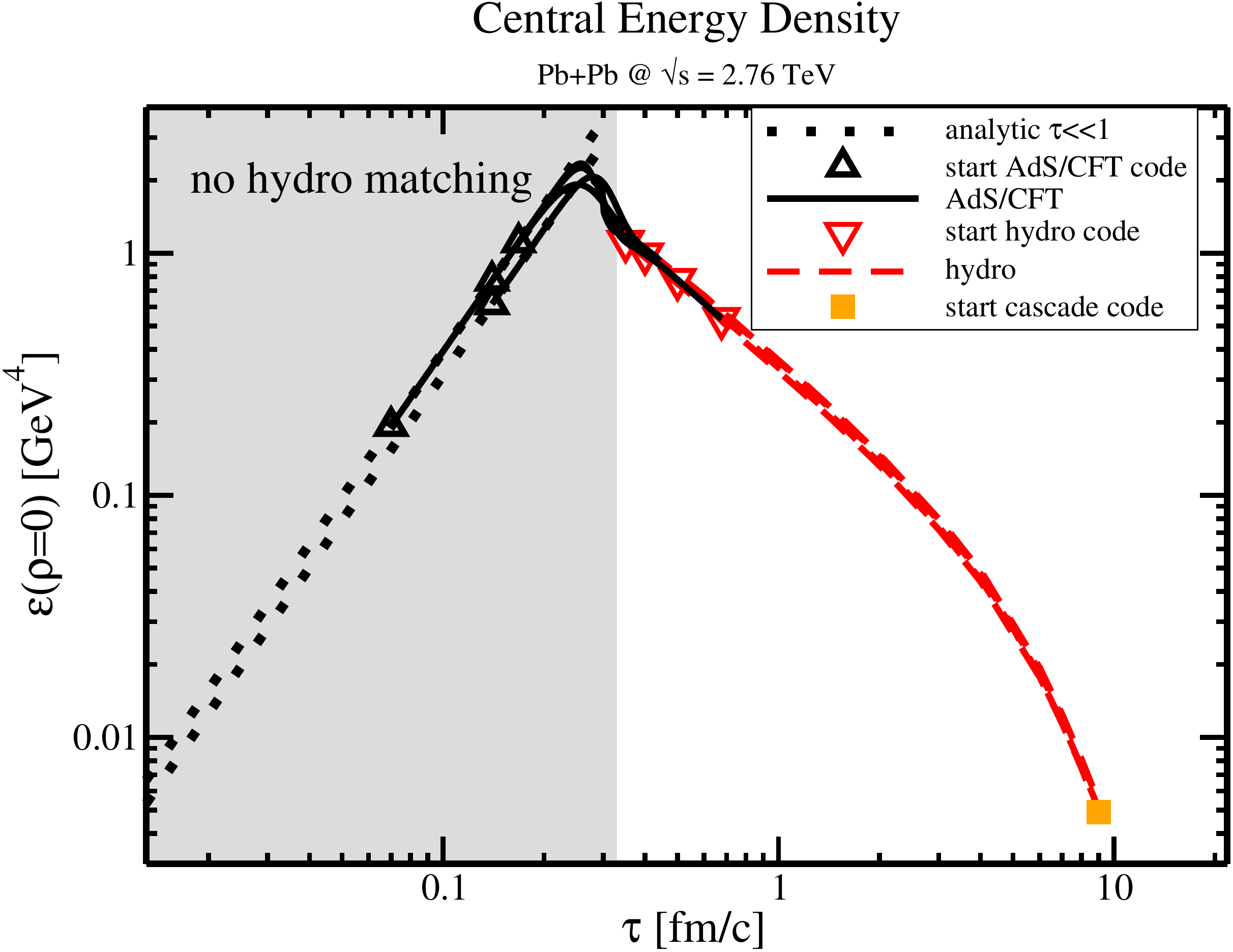}
\caption{
Time evolution of the energy density at the center of the fireball for different values of the regulator $\sigma$, different AdS/CFT starting times $\tau_{\rm init}$ and different AdS/hydro switching times $\tau_{\rm hydro}$. Shown are the analytic early time result (dotted), the numerical AdS/CFT evolution (full lines), the numerical hydro evolution (dashed lines) and the conversion point to the hadron cascade. For $\tau\lesssim 0.35$ fm/c, no sensible matching from AdS/CFT to a hydrodynamic evolution is possible (``no hydro matching''), cf.~Fig.~\ref{fig:aniso1}.}
\label{fig:ed1}
\end{figure}

\begin{figure}[t]
\includegraphics[width=\linewidth]{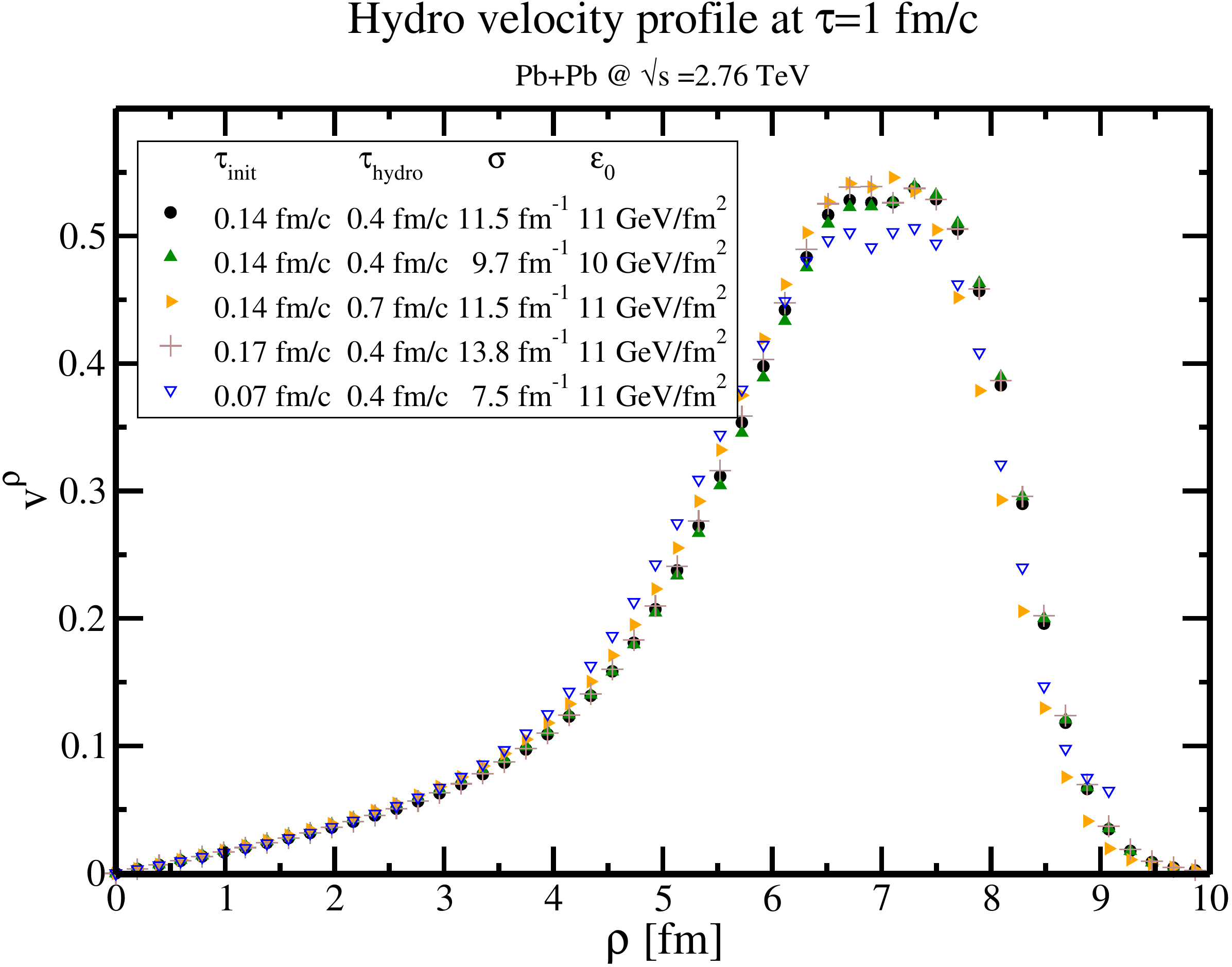}
\caption{Radial velocity profile at $\tau=1$ fm/c for different AdS/CFT starting times $\tau_{\rm init}$, different AdS/hydro switching times $\tau_{\rm hydro}>0.35$ and different values for the regulator $\sigma$ (in a.u.). One observes that when normalized to the same final multiplicity, all these choices lead to similar 
velocity profiles.}
\label{fig:vel1}
\end{figure}

\begin{figure}[t]
\includegraphics[width=\linewidth]{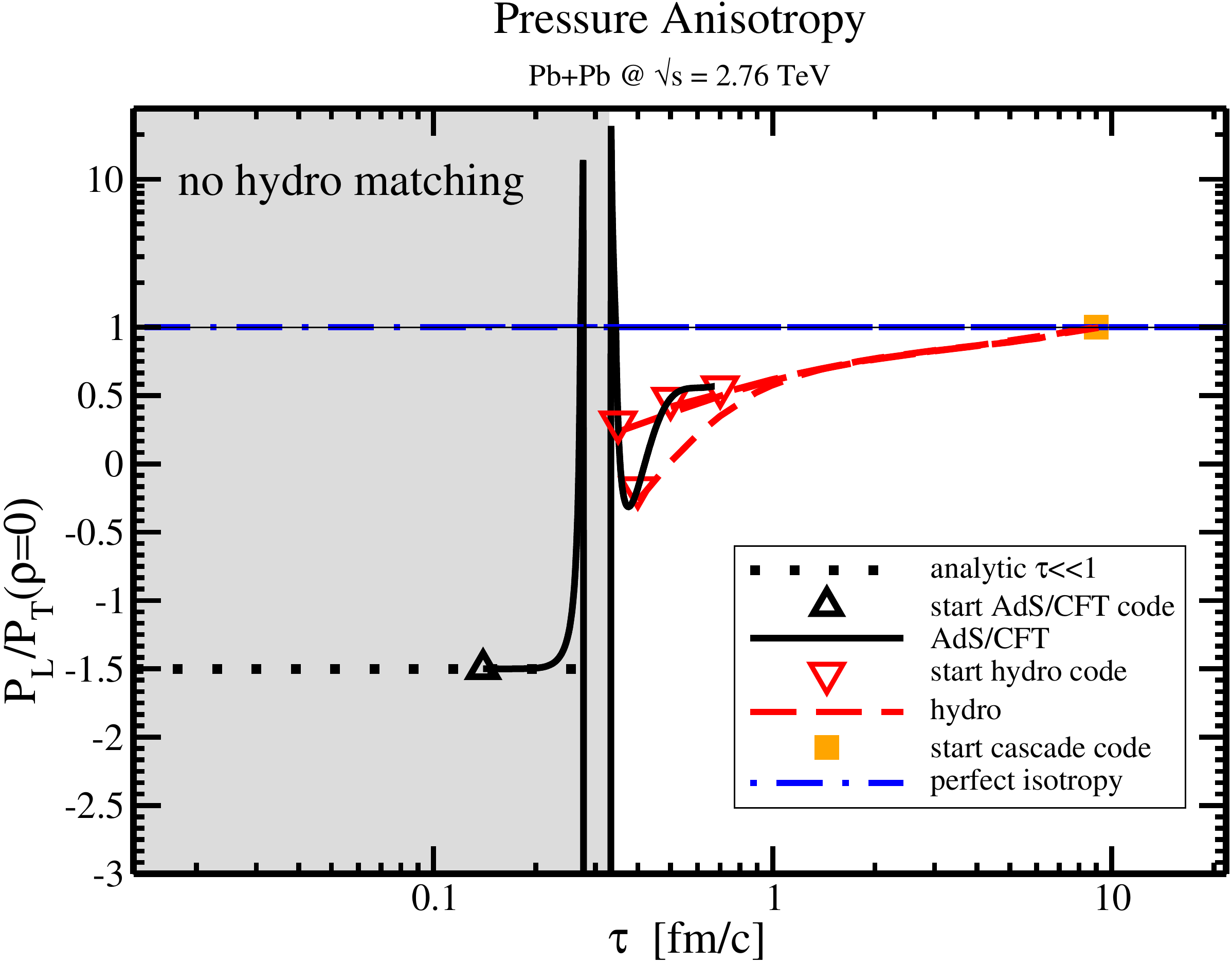}
\caption{Time evolution of the pressure anisotropy $P_L/P_T$ at the center of the fireball for single values of $\sigma,\tau_{\rm init}$ but multiple AdS/hydro switching times $\tau_{\rm hydro}$. For $\tau\lesssim 0.35$ fm/c, the pressure anisotropy is wildly varying, prohibiting a sensible matching to hydrodynamics. At later times, matching to hydrodynamics can be performed (indicated by triangles down) and leads to approximately universal late-time evolution until freeze-out to the hadron cascade (indicated by square). }
\label{fig:aniso1}
\end{figure}

\begin{figure}[t]
\includegraphics[width=\linewidth]{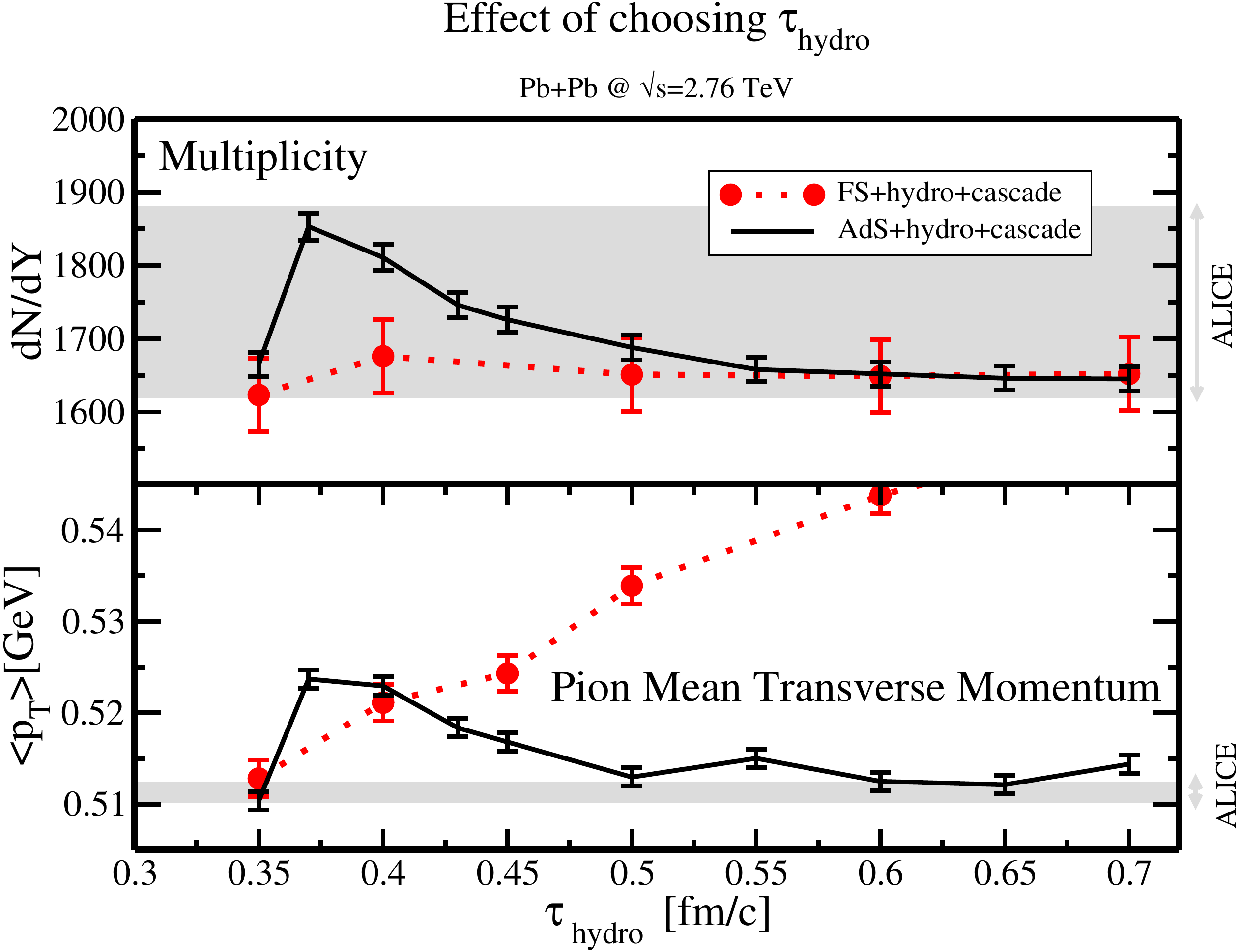}
\caption{Final $dN/dY$ and pion $\langle p_T\rangle$ as a function of the hydro switching time $\tau_{\rm hydro}$ for a single value of $\sigma,\tau_{\rm init}$ (AdS+hydro+cascade) compared to experimental data (``ALICE''). AdS results seem to be independent of $\tau_{\rm hydro}$ provided that $\tau_{\rm hydro}>0.5$ fm/c. By contrast, results for FS models with $\tau_{\rm init}=0.05$ fm/c exhibit strong $\tau_{\rm hydro}$ dependence. Error bars correspond to accumulated numerical error.} 
\label{fig:multi}
\end{figure}

\begin{figure}[t]
\includegraphics[width=\linewidth]{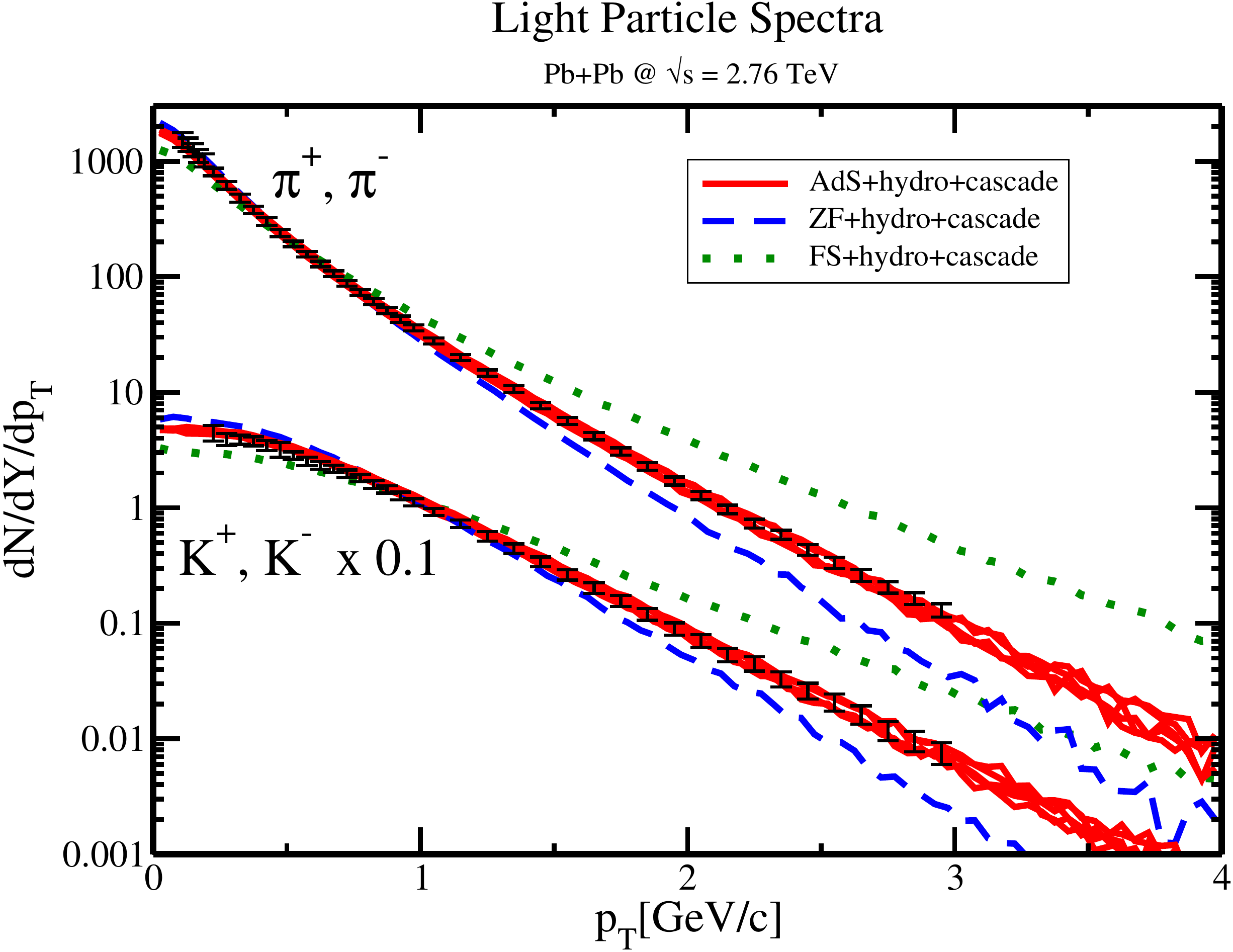}
\caption{Pion and Kaon momentum spectra for 0-5\% most central $Pb+Pb$ collisions at $\sqrt{s}=2.76$ TeV. Experimental measurements (ALICE, Ref.~\cite{Abelev:2012wca}) are compared to our AdS+hydro+cascade model (lines correspond to different choices of $\epsilon_0,\sigma$), the ZF and FS model initialized from Eq.~(\ref{Fermi}) followed by hydro at $\tau_{\rm hydro}=1$ fm/c.
}
\label{fig:spectra}
\end{figure}

\noindent
{{\bf 3. Results.}}
Matching our numerical relativity, viscous hydrodynamics and hadron cascade simulations onto one another we obtain the time-evolution of the energy density for $Pb-Pb$ collisions at $\sqrt{s}=2.76$ TeV (see Fig.~\ref{fig:Temp1}). The results depend on our choices of $\epsilon_0,\tau_{\rm init}$ and $\sigma$, which are all parameters that in principle could be fixed by a more complete calculation. Requiring that for constant $\tau_{\rm init}$ and $\sigma$ our $dN/dY$ matches the experimental value fixes $\epsilon_0$. Different combinations of $\tau_{\rm init}$ and $\sigma$ will have similar late-time energy densities (cf. Fig.~\ref{fig:ed1}), but originate from different early-time histories  and the pre-equilibrium evolution reported in Figs.~\ref{fig:ed1},\ref{fig:aniso1} should be considered uncertain.  However, we find that for fixed $dN/dY$
also the late time radial flow velocity and final light hadron spectra are essentially unaffected by our choice of $\tau_{\rm init}$ or $\sigma$ (see Figs.~\ref{fig:ed1}--\ref{fig:spectra}). 

This is evident when comparing the resulting hydrodynamic radial velocity at $\tau=1$ fm/c shown in Fig.~\ref{fig:vel1}. Different values for $\tau_{\rm init},\tau_{\rm hydro}$ and $\sigma$ collapse onto an approximately universal velocity profile. Because the subsequent evolution follows hydrodynamics, this is also true for the velocity profile for all later times. We therefore expect our late time results to be robust.

One is not completely free in specifying $\tau_{\rm init}$ or $\sigma$. The coordinate singularity at $\tau=0$ prevents going to very early times, while one naturally has to start the AdS/CFT code long before the time hydrodynamics is expected to be applicable. In practice we found $0.07\leq \tau_{\rm init}(\rm fm)\leq 0.17$ to be a good range. For $\sigma$ one has to make sure the AdS spacetime is sufficiently regular to allow for a stable evolution. 
In practice, we found $7.5 \leq \sigma ({\rm fm}^{-1}) \leq 14$ to work well.


The time evolution of the pressure anisotropy shown in Fig.~\ref{fig:aniso1} indicates
a strongly varying, and occasionally negative, longitudinal pressure prohibiting any hydrodynamic description for $\tau<0.35$ fm/c. Besides the strongly varying anisotropy, 
we also typically encounter a closed region in space-time where the system is so far from equilibrium that a local rest frame does not seem to exist, which we plan to report on in future work.
In principle, one could choose any value of $\tau_{\rm hydro}>0.35$ fm/c; however, switching at very late times $\tau_{\rm hydro}\gg1$ fm/c is not recommended because of  the prohibitive computational cost of the  numerical relativity code 
and the fact that at later times the system has cooled down to temperatures where the QCD EoS is no longer close to the conformal EoS in the AdS/CFT code. For $\tau>0.35$ fm/c we can attempt to match the pre-equilibrium phase onto viscous hydrodynamics at $\tau=\tau_{\rm hydro}$, which surprisingly seems to lead to roughly similar final results even when $P_L\simeq 0$  (cf. Fig.~\ref{fig:aniso1}). A more refined result can be gained by considering the dependence of the final light hadron spectra on the choice $\tau_{\rm hydro}$ discussed below.

\noindent
{{\bf 4. Resulting particle spectra 
}}

In order to compare our thermalizing strongly coupled model we have considered two other (extreme) possibilities for the initial stage before $\tau_{hydro}$. The first has $P_L = 0$, which gives zero coupling boost-invariant free streaming (FS), whereas the second has $P_T=0$, which hence has zero pre-equilibrium radial flow (ZF). These models never lead to thermalization, but operationally one can switch to hydrodynamics at some time $\tau_{\rm hydro}$.

Fig. \ref{fig:multi} shows the dependence of the final multiplicity and pion mean transverse momentum on the hydro switching time $\tau_{\rm hydro}$ for the AdS and FS models. 
For the final stage hadron cascade only hydro information for $\tau>1$ fm/c is used. Fig. \ref{fig:multi} indicates that final $dN/dY,\langle p_T\rangle$ in our AdS model are constant, provided one switches to hydrodynamics after the far-from-equilibrium regime has ended (at about $\tau \approx 0.5$ fm/c). This suggests that our model reaches hydrodynamics dynamically and hence results are insensitive to the choice of $\tau_{\rm hydro}$. 
%
In contrast, for the FS and ZF models $\langle p_T\rangle$ depends on $\tau_{hydro}$, which hence only reproduces the data for a specific choice. Thus,
while not ruled out by data, the FS and ZF models are less predictive than the AdS model. 

However, Fig.~\ref{fig:multi} demonstrates that FS leads to radial flow.
Hence findings of radial flow in $p+Pb$ and $d+Au$ systems \cite{Abelev:2013haa,Abelev:2008ab,Sickles:2013yna} 
may not necessarily indicate strongly coupled matter. To discriminate between weak and strong coupling
one may use angular correlations which are not built up in FS \cite{Kolb:2003dz}.



In Fig.~\ref{fig:spectra} we show the results for the final pion and kaon transverse momentum spectra in comparison to data for central $Pb+Pb$ collisions at $\sqrt{s}=2.76$ TeV from the ALICE experiment \cite{Abelev:2012wca}. The integral over the momentum spectra corresponds to the total multiplicity which we fixed by hand. However, Fig.~\ref{fig:spectra} shows that our AdS+hydro+cascade model matches the shape of the experimental data almost perfectly up to the highest transverse energies measured, independent of our choices for $\tau_{\rm init}, \tau_{\rm hydro}, \sigma$.

\noindent
{{\bf 5. Conclusions and discussion.}}

In this work we have presented the first fully dynamical multi-physics simulation of central nuclear collision at LHC energies. This simulation includes a simulation of the equilibration of the bulk of the system using the AdS/CFT correspondence. When normalized to the same multiplicity, our framework is approximately insensitive to the AdS initialization time $\tau_{\rm init}$, the choice of bulk parameter $\sigma$ and the AdS/hydro switching time $\tau_{\rm hydro}$, provided the switching occurs later than $\sim 0.5$ fm/c. 
This is in contrast to non-thermalizing models such as FS+hydro+cascade where results depend on choices for $\tau_{\rm init},\tau_{\rm hydro}$.

Because of the dynamical treatment of the pre-equilibrium stage and the insensitivity to our free parameters, our model is more constrained than a standard hydro+cascade model. In particular, we find that the transverse pressure is consistently higher than the longitudinal pressure, during most of the evolution (Fig.~\ref{fig:aniso1}). Very encouragingly, the model  turns out to have light particle spectra in excellent agreement with experimental data for $Pb+Pb$ collisions at $\sqrt{s}=2.76$ TeV.

We regard this work as the first step towards a truly realistic simulation of high energy nuclear collisions. Many aspects of our work can and should be improved in future work. For instance, we plan to do away with the bulk parameter $\sigma$ by simulating the full shock-wave collision process (cf.~\cite{Casalderrey-Solana:2013aba}) and simulate event-by-event non-central collisions by employing recent development in black-hole evolution schemes \cite{Bantilan:2012vu}. 


\noindent
{{\bf Acknowledgments.}}
This work was supported by the U.S. Department of Energy Office of Science through grant number DE-FG02-03ER41259 and award No. DE-SC0008132 as well as a Utrecht University Foundations of Science grant. We thank T.~Peitzmann and L.~Yaffe for discussions.

\vspace{-0.5cm}
\bibliographystyle{plain}

\begin{thebibliography}{99}
\vspace{-0.4cm}
\bibitem{Abelev:2012cya}
  B.~Abelev {\it et al.}  [ ALICE Collaboration],
  arXiv:1212.2001 [nucl-ex].


\bibitem{Adamczyk:2012ku}
  L.~Adamczyk {\it et al.}  [STAR Collaboration],
  Phys.\ Rev.\ C {\bf 86} (2012) 054908
  [arXiv:1206.5528 [nucl-ex]]. 

\bibitem{Alver:2008aq}
  B.~Alver, M.~Baker, C.~Loizides and P.~Steinberg,
  arXiv:0805.4411 [nucl-ex].


\bibitem{deHaro:2000xn}
 S.~de Haro, S.~N.~Solodukhin and K.~Skenderis,
  Commun.\ Math.\ Phys.\  {\bf 217}, 595 (2001).

\bibitem{Romatschke:2013re}
  P.~Romatschke and J.~D.~Hogg,
  JHEP {\bf 1304} (2013) 048
  [arXiv:1301.2635 [hep-th]].

\bibitem{Vredevoogd:2008id}
  J.~Vredevoogd and S.~Pratt,
  Phys.\ Rev.\ C {\bf 79} (2009) 044915
  [arXiv:0810.4325 [nucl-th]].

\bibitem{Casalderrey-Solana:2013aba}
  J.~Casalderrey-Solana, M.~P.~Heller, D.~Mateos and W.~van der Schee,
  arXiv:1305.4919 [hep-th].

\bibitem{Grumiller:2008va}
  D.~Grumiller and P.~Romatschke,
  JHEP {\bf 0808} (2008) 027
  [arXiv:0803.3226 [hep-th]].

\bibitem{Taliotis:2010pi}
  A.~Taliotis,
  JHEP {\bf 1009} (2010) 102
  [arXiv:1004.3500 [hep-th]].


\bibitem{Luzum:2008cw}
  M.~Luzum and P.~Romatschke,
  Phys.\ Rev.\ C {\bf 78} (2008) 034915
   [Erratum-ibid.\ C {\bf 79} (2009) 039903]
  [arXiv:0804.4015 [nucl-th]].


\bibitem{Wu:2011yd}
  B.~Wu and P.~Romatschke,
  Int.\ J.\ Mod.\ Phys.\ C {\bf 22} (2011) 1317
  [arXiv:1108.3715 [hep-th]].


\bibitem{Chesler:2010bi}
  P.~M.~Chesler and L.~G.~Yaffe,
  Phys.\ Rev.\ Lett.\  {\bf 106}, 021601 (2011) 
  [arXiv:1011.3562 [hep-th]].

\bibitem{vanderSchee:2012qj}
  W.~van der Schee,
  Phys.\ Rev.\ D {\bf 87} (2013) 061901
  [arXiv:1211.2218 [hep-th]].

\bibitem{Chesler:2013aa}
  P.M.~Chesler, L.G.~Yaffe,
  [arXiv:1309.1439 [hep-th]].

\bibitem{Pratt:2010jt}
  S.~Pratt and G.~Torrieri,
  Phys.\ Rev.\ C {\bf 82} (2010) 044901
  [arXiv:1003.0413 [nucl-th]].

\bibitem{Novak:2013bqa}
  J.~Novak, K.~Novak, S.~Pratt, C.~Coleman-Smith and R.~Wolpert,
  arXiv:1303.5769 [nucl-th].


\bibitem{Abelev:2013haa}
  B.~B.~Abelev {\it et al.}  [ALICE Collaboration],
  arXiv:1307.6796 [nucl-ex].

\bibitem{Abelev:2008ab}
  B.~I.~Abelev {\it et al.}  [STAR Collaboration],
  Phys.\ Rev.\ C {\bf 79} (2009) 034909
  [arXiv:0808.2041 [nucl-ex]].

\bibitem{Sickles:2013yna}
  A.~M.~Sickles,
  arXiv:1309.6924 [nucl-th].
\bibitem{Kolb:2003dz}
  P.~F.~Kolb and U.~W.~Heinz,
  In *Hwa, R.C. (ed.) et al.: Quark gluon plasma* 634-714
  [nucl-th/0305084].

\bibitem{Abelev:2012wca}
  B.~Abelev {\it et al.}  [ALICE Collaboration],
  Phys.\ Rev.\ Lett.\  {\bf 109} (2012) 252301
  [arXiv:1208.1974 [hep-ex]].






\bibitem{Bantilan:2012vu}
  H.~Bantilan, F.~Pretorius and S.~S.~Gubser,
  Phys.\ Rev.\ D {\bf 85} (2012) 084038
  [arXiv:1201.2132 [hep-th]].

\bibitem{footnote1}
As in previous work we had to introduce a regulator energy density to stabilize the code for regions with little energy. We checked that this regulator does not affect physical results presented here.


\end{thebibliography}

\end{document}